\documentclass[aps,prl,reprint,english,superscriptaddress, floatfix]{revtex4-1}
\usepackage{amsthm}
\usepackage{amssymb}
\usepackage{amsmath}
\usepackage{epsf}
\usepackage[scanall]{psfrag}
\usepackage{dsfont}
\usepackage[T1]{fontenc}
\usepackage[latin9]{inputenc}
\usepackage{comment}
\usepackage{lmodern}
\usepackage[scanall]{psfrag}
\usepackage{geometry}
\usepackage{comment}

\usepackage{caption,subcaption}
\usepackage{amsmath}
\usepackage{amssymb}
\usepackage{graphicx}
\usepackage[export]{adjustbox}
\usepackage[usenames,dvipsnames]{color}
\usepackage{tikz}
\usepackage{pgfplots}
\usepackage{multirow}
\usepackage{placeins}
\usepackage{appendix}

\usepackage{babel}

\begin{document}

\title{Kinetic Transition Networks for the Thomson Problem and Smale's 7th Problem}

\author{Dhagash Mehta}
\email{dmehta@nd.edu}
\affiliation{Department of Applied and Computational Mathematics and Statistics, University of Notre Dame, 
Notre Dame, IN 46556, USA.}
\affiliation{Department of Chemical and Biomolecular Engineering, University of Notre Dame, 
Notre Dame, IN 46556, USA.}
\affiliation{Centre for the Subatomic Structure of Matter, Department of Physics, School of Physical Sciences,
University of Adelaide, Adelaide, South Australia 5005, Australia.}

\author{Jianxu Chen}
\email{jchen16@nd.edu}
\affiliation{Department of Computer Science and Engineering, University of Notre Dame, Notre Dame, IN 46556, USA.}

\author{Danny Z. Chen}
\email{dchen@nd.edu}
\affiliation{Department of Computer Science and Engineering, University of Notre Dame, Notre Dame, IN 46556, USA.}

\author{Halim Kusumaatmaja}
\email{halim.kusumaatmaja@durham.ac.uk}
\affiliation{Department of Physics, Durham University,
South Road, Durham DH1 3LE, United Kingdom.}

\author{David J.~Wales}
\email{dw34@cam.ac.uk}
\affiliation{University Chemical Laboratories, Lensfield Road, Cambridge CB2 1EW, United Kingdom.}

\begin{abstract}
\noindent 
The Thomson Problem, arrangement of identical charges on the surface of a sphere, has 
found many applications in physics, chemistry and biology. Here we show that the energy landscape of 
the Thomson Problem for $N$ particles with $N=132,\ 135,\ 138,\ 141,\ 144,\ 147,$ and 150 is single funnelled, 
characteristic of a structure-seeking organisation where the global minimum is easily accessible.
Algorithmically constructing starting points close to the global minimum 
of such a potential with spherical constraints is one of Smale's 18 unsolved problems in mathematics for the 21st 
century because it is important in the solution of univariate 
and bivariate random polynomial equations. By analysing the kinetic transition networks, 
we show that a randomly chosen minimum is in fact always `close' to the global minimum 
in terms of the number of transition states that separate them, a characteristic of small world networks.
\end{abstract} 

\maketitle

\section{Introduction}

The Thomson Problem is an important model in both physics and chemistry and is easily stated: find the
minimum energy of a system composed of $N$ identical charges that are constrained to move
on the surface of a sphere of unit radius, i.e.~minimise the potential energy function
\begin{equation}
V_{\mbox{Th}}(\textbf{r}) =\sum_{1\leq i<j\leq N}\frac{1}{r_{ij}},
\end{equation}
where $r_{ij}=\sqrt{(x_{i}-x_{j})^{2}+(y_{i}-y_{j})^{2}+(z_{i}-z_{j})^{2}}$, 
subject to spherical constraints $x_{i}^{2}+y_{i}^{2}+z_{i}^{2}=1$ for all $i = 1,\dots, N$.

The model was proposed by J.~J.~Thomson as a representation of atomic
structure \cite{thomson1904xxiv}, although it was soon abandoned in this context.
However, it has subsequently found many applications
in physics, chemistry and biophysics because it
captures the competition between local order for
neighbouring particles and long-range constraints due to the curvature and
geometry. In particular, it is not generally possible to arrange every particle in an
identical environment. 
The resulting model can provide insight into the forces governing far more complex systems,
such as the arrangement of
proteins in shells of spherical viruses \cite{caspar1962physical,marzec1993pattern,bruinsma2003viral,zandi2004origin},
fullerene patterns for carbon clusters \cite{kroto1985c},
the surface ordering of liquid metal drops confined in Pauli traps \cite{davis1997history},
and the behaviour of colloidal particles trapped at a fluid-fluid 
interface \cite{disc03,LipowskyBMNB05,EinertLSBB05,Kusumaatmaja2013,meng2014elastic,IrvineVC10,IrvineBC12,PhysRevE.88.012405}.

In the mathematics and computer science communities, the Thomson Problem has gained special attention
because it appears in the 7th problem in Steven Smale's list of eighteen
unsolved problems for the 21st century \cite{smale1998mathematical,smale1998mathematical-2}.
The problem as posed by Smale \footnote{Smale's 7th problem rather 
considers a general form of the potential, $r^{\alpha}$, with the spherical constraints. 
The special values $\alpha = 0$ and $1$ correspond to a logarithmic potential and the Thomson Problem, respectively.
In
Ref.~\cite{smale1998mathematical-2} $\alpha=0$ is considered as the `main' problem (and the Thomson Problem 
listed as a special case of the general problem) and more mathematicians are focused on 
this version of the problem. 
$\alpha=1$ is a popular model in chemistry and physics.
In the asymptotic limit, both versions of the problems are similar in the following sense:
a solution to Thomson Problem is known to have an asymptotically optimal logarithmic potential. This statement is proven both ways 
in Refs.~\cite{leopardi2013discrepancy,beltran2015facility}. In any case, both versions of the problem share the same importance and 
qualitative features.} requires us
to algorithmically \cite{blum2012complexity} construct a collection of starting points, 
say $\textbf{r}^*$, so that for a given number of particles
$N$, $V_{\mbox{Th}}(\textbf{r}^*) - V_{\mbox{Th}}(\textbf{r}_{\mbox{gm}}) \leq \mbox{const} \times \log N$, for $N\geq 2$.
Here, $\textbf{r}_{\rm gm}$ is the global minimum for the given $N$. 

Constructing such points corresponds to finding a good 
starting system of polynomial equations to locate all the complex solutions for sets of polynomial 
equations and to realise
the Fundamental Theorem of Algebra \cite{shub1993complexity,armentano2011minimizing}. 

Given its broad relevance, the global minimum of the Thomson Problem has been studied 
extensively \cite{siman86,ErberH91,constrained,ErberH95,altschuler97,%
ErberH97,genetic,altschuler05,AltschulerP06,WalesU06,WalesMA09,brauchart2009riesz,brauchart2012next,Bowick}.
Analytical solutions are known for $N=2-6$ and $12$ \cite{andreev1996extremal,schwartz20105}.
For $N=4, 6$ and $12$, they correspond to Platonic solids.
The global minimum structures for larger $N$ have been tackled computationally
\cite{saff1997distributing,kuijlaars1998asymptotics,
disloc99,livshits99,BowickNT00,BowickCNT02,BowickCNT06,AltschulerP06,WalesU06,PhysRevE.78.010601,lafave2014discrete}.
Most of these studies, however, only focus on the global minima.
In the present work we search extensively for (most, if not all) the minima and transition states of the
Thomson Problem at some selected sizes up to  $N = 150$, and study the networks defined by these minima
and the transition states that connect them. Using disconnectivity graphs \cite{beckerk97,walesmw98},
we find that the Thomson Problem
exhibits a single funnel for all $N$ considered. We also show that the networks
exhibit typical
characteristics of a small world network. 
Hence, in the context
of Smale's 7th problem, all minima are only a few steps from the global minimum.

\section{Potential Energy Landscapes of the Thomson Problem}
We begin by collecting some basic definitions: a local minimum of a potential is a point in the configuration space where
the gradient vanishes and
at which the Hessian matrix has no negative eigenvalues.
The minimum at which the potential attains the lowest value is the global minimum. 
A transition state \cite{murrelll68} is a configuration where the gradient vanishes, 
and exactly one eigenvalue of the Hessian matrix 
is negative. Transition states define connections between pairs of minima via steepest-descent paths.
Hence, we can construct kinetic transition networks \cite{NoeF08,pradag09,Wales10a} of local minima and
the transition states that link them for a given potential.

To perform an extensive search for local minima and transition states 
of the Thomson Problem, 
we employed the GMIN \cite{gmin} and OPTIM
programs \cite{optim}.
To identify likely global minima we applied basin-hopping global optimisation
\cite{lis87,walesd97a,waless99,Wales03}. In this method, random geometrical
perturbations are followed by energy minimisation, and moves are accepted or
rejected based upon the energy differences between local minima. This
procedure transforms the energy landscape system into the
set of catchment basins for the local minima.
For all the
minimisations in the present work a modified version of the limited-memory
Broyden--Fletcher--Goldfarb--Shanno (LBFGS) algorithm  \cite{Nocedal80,lbfgs}
was used. This scheme has proved to be the most efficient in recent
benchmarks \cite{AsenjoSWF13}.

We used a combination of the doubly-nudged elastic band (DNEB) and hybrid
eigenvector-following techniques to find the transition states
\cite{TrygubenkoW04}. In DNEB, a series of images interpolate between the two
end points, and the total energy is minimised subject to spring constraints between
adjacent images.  Maxima in the DNEB path are then adopted as transition state
candidates, which we refine to high precision using hybrid
eigenvector-following \cite{HenkelmanUJ00,munrow99,kumedamw01}. 
Here we use a Rayleigh-Ritz approach \cite{Wales03} to compute the smallest non-zero eigenvalue, 
and take an uphill step along the corresponding eigendirection. We then
minimize in the tangent space for a limited number of steps, so that the
gradient does not develop a significant component in the uphill direction. We
project out all components corresponding to  zero Hessian eigenvalues. For the
Thomson Problem these are the three modes associated with overall rotations
around the x, y, and z axes. Using 
spherical polar coordinates $(\phi,\theta)$, the corresponding
eigenvectors are $\hat{e}_x = (\cos{\theta}\cos{\phi}/\sin{\theta}, \sin{\phi})$,
$\hat{e}_y = (-\cos{\theta}\sin{\phi}/\sin{\theta}, \cos{\phi})$ and
$\hat{e}_z = (1, 0)$.

For each transition state we applied small displacements in the two
downhill directions and minimised the energy to identify
minimum-transition state-minimum triplets. In some cases, new local minima
may be identified in this procedure. By iterating the process, we
systematically build a connected database of stationary points.

\begin{figure}
    \centering
    \includegraphics[width=0.45\textwidth]{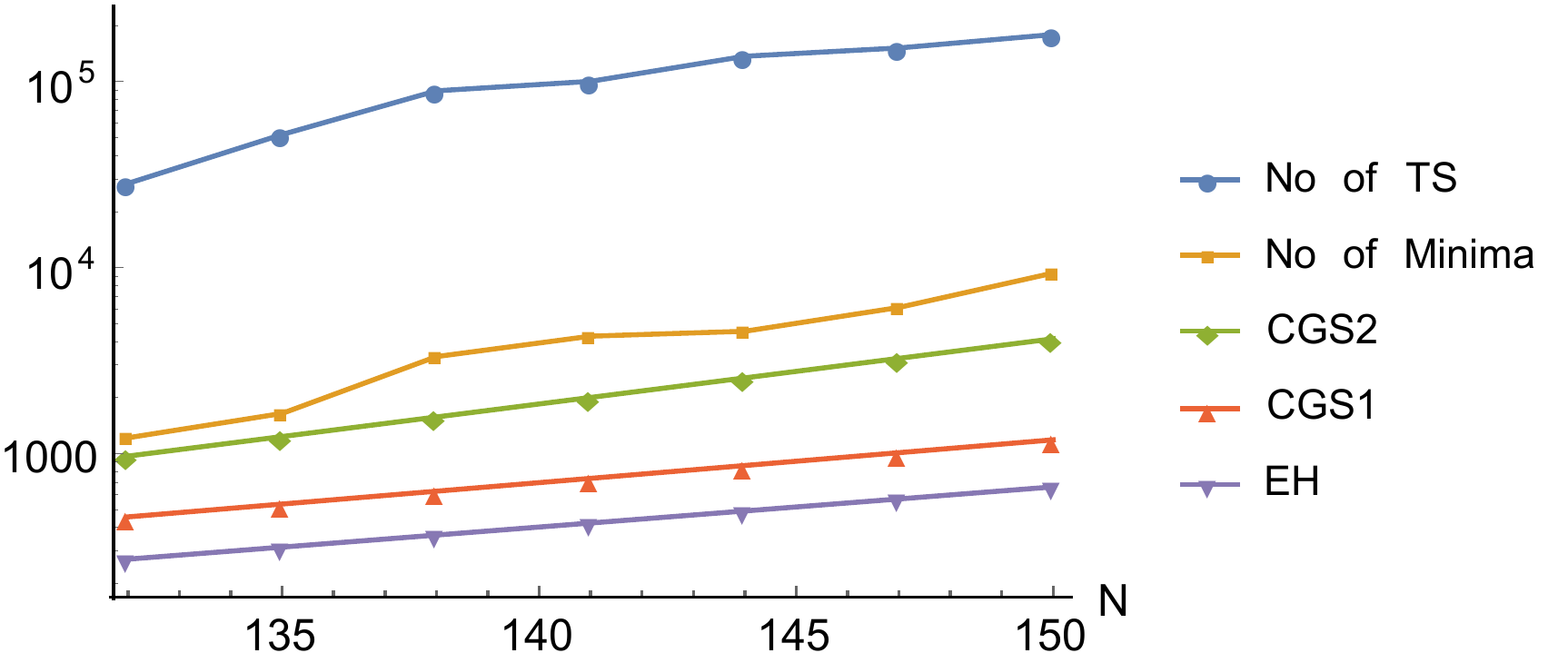}
    \caption{`TS' refers to `Energetically Distinct Transition Sates', `EH' refers to the estimate of the number of minima 
    given in \cite{ErberH95}, `CGS1' is a recent \cite{Calef:2015} fit to the number of minima found in
    previous calculations, 
    and `CGS2' is an estimate \cite{Calef:2015} 
    for the number of minima (we use the maximum out of those suggested in \cite{Calef:2015}.) The lines connecting data points 
    are a guide to the eye.}
    \label{plotNvsMin}
\end{figure}

Our results for the number of minima and transition states at $N=132, 135, 138, 141, 144, 147$ and $150$ 
are shown in Figure \ref{plotNvsMin}. In all cases \footnote{We compare our results with the 
estimate on the number of minima, $0.382 \exp(0.0497 N)$,
as given in \cite{ErberH95}, the number of minima from a fit to the number of minima found,
$(0.31701 \pm 0.1) \exp( (0.0518 \pm 0.0012) N )$ and estimated number of minima 
$\exp(-3.97635 \pm 0.1992) \exp(0.0789298 \pm 0.00176) N )$
(with choosing the '$+$' sign for both '$\pm$' in our comparison), 
as given in \cite{Calef:2015}.} 
we have obtained more minima than previous results
given in the literature \cite{ErberH95,Calef:2015}.
It is known that usually the number of local minima for a molecular system
increases exponentially with system size \cite{stillingerw84,ErberH95,WalesD03}.
This scaling applies here, and locating likely candidates for the global minimum becomes increasingly
difficult for larger systems. However, the rate of increase is relatively slow for the Thomson Problem,
because the Coulomb potential is long-ranged \cite{doyewb95,doyew96d,doyew97a}.

Disconnectivity graphs provide a
powerful way to visualize the organization of minima and transition states \cite{beckerk97,walesmw98}. 
A disconnectivity graph is a tree graph where the vertical axis corresponds to 
the potential energy. Each line terminates at the energy of a local minimum, and the minima are joined at the lowest energy 
where they can interconvert for regularly spaced energy thresholds. These connections are defined by the highest transition state 
on the lowest energy path between each pair of minima.

\begin{figure}[h!]
    \centering
    \includegraphics[width=0.33\textwidth]{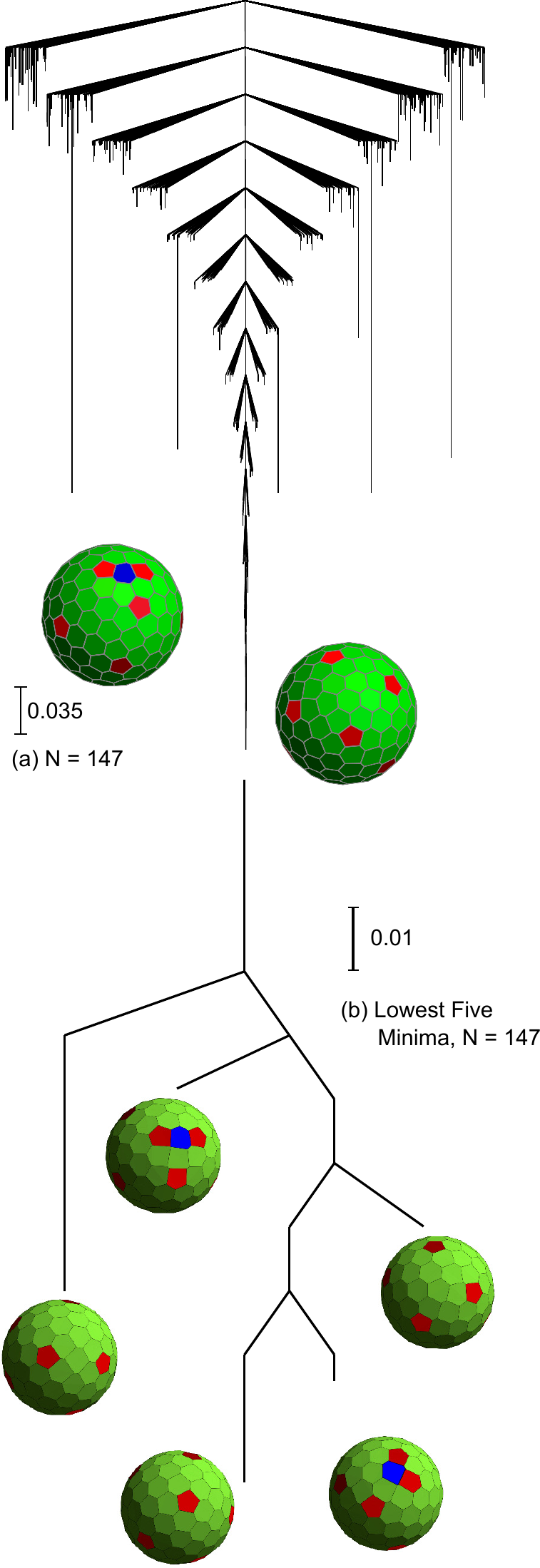}
    \caption{(a) The disconnectivity graph for $N=147$. 
    In all cases considered here, the disconnectivity graphs show a typical
    structure-seeking `palm tree' organisation.
    In panel (b) The insets show the structures of the lowest five minimum energy configurations. 
    The pentagons, hexagons, and heptagons
    indicate that the charges in the centre of the polygons have 5, 6 and 7
    neighbours respectively.}
    \label{DisGraph}
\end{figure}

The disconnectivity graph for the databases considered here is shown in Figure \ref{DisGraph} for $N = 147$. 
It shows a typical structure-seeking `palm tree' organisation \cite{walesmw98},
where there is only a single funnel 
in the potential energy landscape and the energy barriers separating local energy minima and the global 
minimum are small. The same qualitative features are observed for all particle numbers we have considered, up to $N=150$. 
This organisation is associated with efficient relaxation to the global minimum.

\section{Network Properties}


For each $N$, we construct a network $G$ in which each node corresponds to a unique minimum
energy structure,
where permutation-inversion isomers are lumped together. 
Hence, there are $\mathcal{N}_{\mbox{min}}$ nodes.
Two nodes are connected by an edge if a 
transition state exists between the corresponding minima.
For each $N$ we have obtained $\mathcal{N}_{\mbox{ts}}$ energetically distinct transition states.
Here we need only consider whether two minima 
are connected or not, and more than one distinct transition state between a pair of minima only counts as a single edge.
We call $\mathcal{N}_{\mbox{edge}}$ the number of edges for the given network, with 
$\mathcal{N}_{\mbox{ts}} \ge \mathcal{N}_{\mbox{edge}}$. 
As in previous work for Lennard-Jones clusters \cite{Doye02,doye2005characterizing,massen2005identifying},
our networks are undirected and unweighted graphs, and hence 
agnostic about all other information, such as 
barrier heights or transition rates between the minima.

To further describe our findings, we first define two quantities.
The characteristic path length ($L$) is the 
average of the shortest path between each pair of nodes in $G$. 
We apply an all-pairs shortest path algorithm \cite{cormen2009introduction} to 
the graph $G$ to compute the distance between every two nodes. $L$ is 
the arithmetic mean of the distances between all pairs of nodes with a finite distance. 
The local clustering coefficient $C_i$ of the $i$th node is defined as the
fraction of pairs of the neighbouring nodes that are connected over all
pairs of the neighbours of the node. The global clustering coefficient $C$ is
the average of $C_i$ for $i = 1, \dots, \mathcal{N}_{\mbox{min}}$.

A graph is a small-world network if the characteristic path length is similar
to, and the clustering coefficient 
is much higher than, the random graph with the same number of nodes and edges \cite{watts1998collective}.
Small-world properties have been observed in a number of networks of dynamic systems. 
Our results for the networks of minima of the Thomson Problem for various values of $N$ 
are reported in Table \ref{tab1}.
The characteristic path length and the clustering coefficient of the corresponding random graph are 
denoted as $L_0$ and $C_0$, respectively. It is evident that $L$ is as small as $L_0$, while $C$ is much larger than $C_0$, and 
hence the small-world characteristics of the networks are established.
The network diameter values, defined in terms of the longest shortest path length, are
5 ($N=132,135,138, 141, 144$), 6 ($N=147$) and 7 ($N=150$).

\begin{figure}
    \centering
    \includegraphics[width=0.4\textwidth]{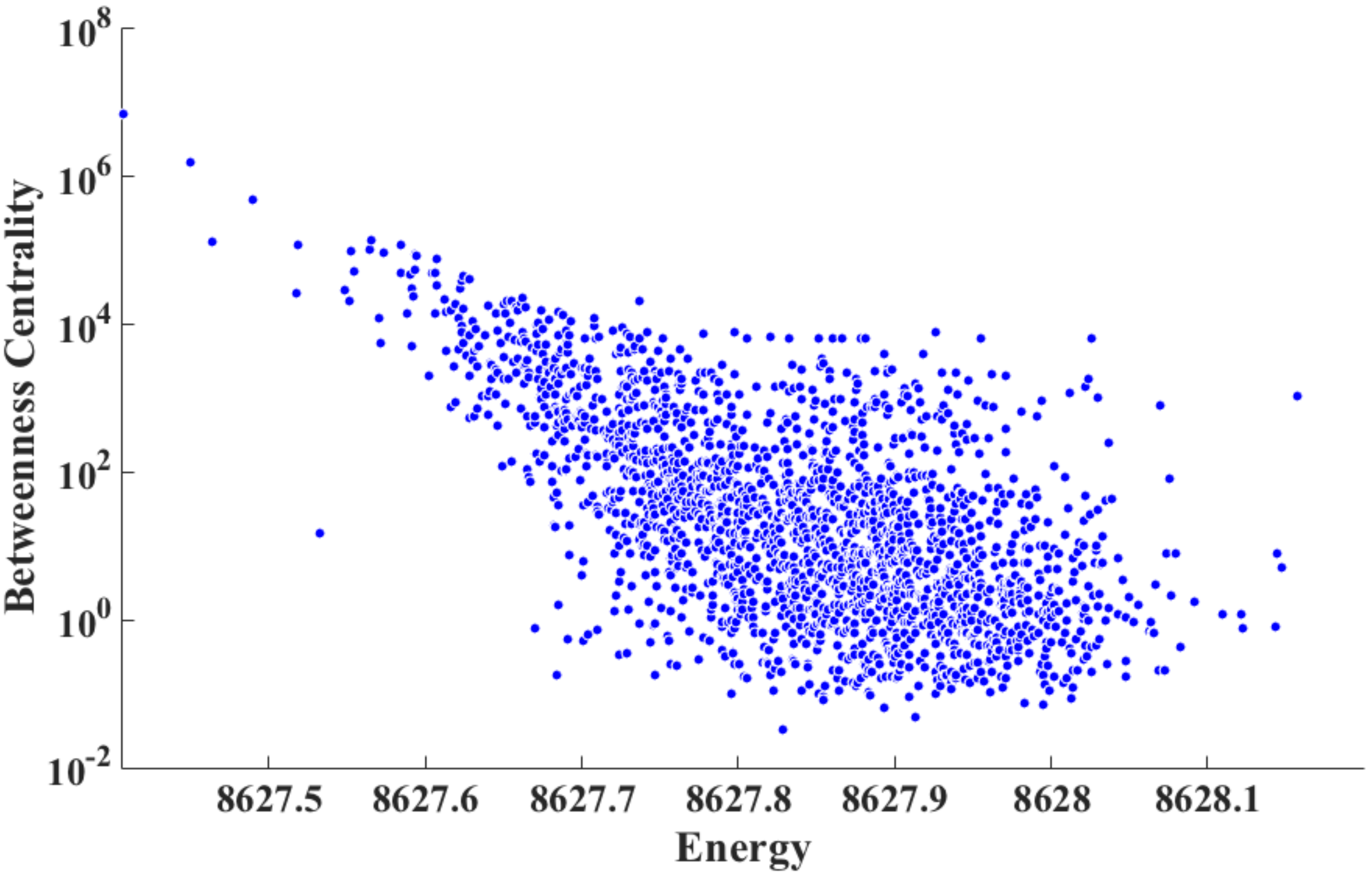}
    \caption{Betweenness centrality of the individual minima for $N=138$. }
    \label{betweenness_centrality}
\end{figure}

\begin{table}[htbp]
   \centering
   \begin{tabular}{@{} | c || c | c | c | c | c | c | c | @{}} 
   \hline
       $N$    & $\mathcal{N}_{\mbox{min}}$ &  $\mathcal{N}_{\mbox{ts}}$ & $\mathcal{N}_{\mbox{edge}}$ & $L$ & $L_0$ & $C$ & $C_0$\\ \hline
132 & 1183 & 28284 & 8700	& 2.3003  & 2.8959 & 0.6032 & 0.0124  \\
135 & 1585 & 51832 & 11285  & 2.2368 & 3.0391 & 0.6932 & 0.009 \\
138 & 3226 & 88999 & 20150 & 2.1791 & 3.4865 & 0.6312 & 0.0039 \\
141 & 4165 & 100085 & 38210 & 2.5684 & 3.1487 & 0.5871 & 0.0044 \\
144 & 4534 & 136519 & 33113 & 2.3369 & 3.4445 & 0.7142 & 0.0032 \\
147 & 6644 & 151299 & 39241 & 2.4314 & 3.8247 & 0.6375 & 0.0018 \\
150 & 9774 & 178728 & 87203 & 2.7601 & 3.5163 & 0.4312 & 0.0018\\
\hline
   \end{tabular}
   \caption{Analysis of networks for different values of $N$. 
   $\mathcal{N}_{\mbox{min}}$, $\mathcal{N}_{\mbox{ts}}$ and $\mathcal{N}_{\mbox{edge}}$  
   are the number of minima (nodes), energetically distinct transition states, and edges, respectively. 
   $L$ and $C$ are the characteristic path length and the clustering coefficient of each network, 
   while $L_0$ and $C_0$ are the characteristic path 
   length and the clustering coefficient of the corresponding random graph.}
   \label{tab1}
\end{table}

\begin{figure}
    \centering
    \includegraphics[width=0.4\textwidth]{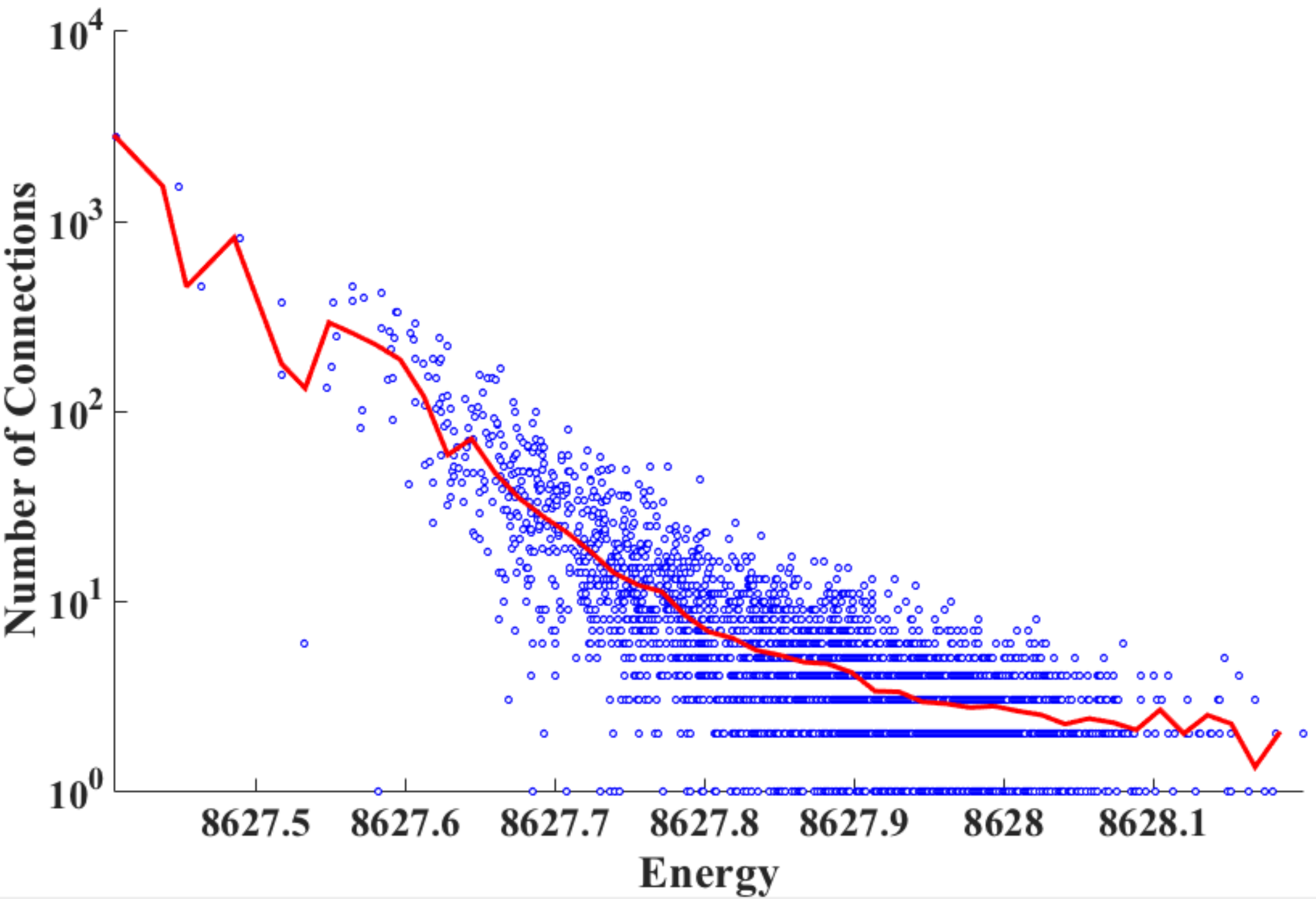}
    \caption{Potential energy minima vs the number of connection at the minimum
is plotted, for $N=138$. The line is the binned average.}
    \label{min_vs_k}
\end{figure}

Another important network property is the betweenness centrality, which is 
defined for each node as the number of shortest paths from 
all nodes to all others that pass through the node under consideration. Betweenness centrality determines
how \textit{central} the node is. In Figure \ref{betweenness_centrality}, we plot the betweenness centrality 
of all the nodes of our networks. Our results reveal that the global minimum of the Thomson Problem 
is always the most central in the network. The number of connections (the node degree) of minima as a function of energy is plotted 
in Figure \ref{min_vs_k}. The figure clearly shows that low-lying minima are highly connected, making them hubs, 
unlike the higher energy minima.

%
%
%
%
%

\section{Discussion and Conclusion}
In this paper we have characterised the potential energy landscape of the
Thomson Problem for selected sizes up to $N = 150$. 
We confirmed that both the number of minima and transition states grow
exponentially with $N$,
albeit with a small exponential factor due to the long range nature of the Coulomb interactions. 
This exponential increase makes the searches for the global minima of the
Thomson Problem progressively more difficult. 
However, by analysing the disconnectivity graphs, we find that 
for the sizes investigated here, the landscapes 
exhibit clear structure-seeking organisation.
The global minima for systems
characterised by such funnelled potential energy landscapes can usually be located easily,
even when the total number of minima is large. 
The asymmetry of the energy barriers 
makes it easy to escape from high to low energy minima, 
while the reverse transitions are significantly slower. 

We have analysed the networks consisting of local minima and the connections between
them defined by transition states, and
find that they exhibit small world properties \cite{watts1998collective}.
Our results provide further evidence to support the conjecture that the
small-world phenomenon might be another generic feature of 
landscapes with the structure-seeking `palm tree' organisation \cite{CarrW08}.
Moreover, we found that the low-lying minima are generally significantly more 
connected than those at higher energy, suggesting
scale-free properties. Further statistical tests \cite{clauset2009power}
are required to confirm this behaviour.
We note that similar kinetic hubs have been identified for biomolecules in
previous work \cite{CavalliHPC03,GfellerdDCR07,CarrW08}.
It is important to note that these networks are static.
Hence, the scale-free phenomenon, if confirmed, requires
explanation beyond the usual preferential attachment schemes \cite{BarabasiA99}.

Our results are relevant for addressing Smale's 7th problem. 
For the sizes considered here, the global minimum is, 
on average, only a few transition states [$O(\log\mathcal{N}_{\mbox{min}})$]
away from any random starting point. 
$\mathcal{N}_{\mbox{min}}$ grows exponentially with $N$, 
and the increase in the average number of connections is linear. 
Since the diameter of the networks is typically 5 or 6 (and at most 7), the
global minimum is never further than 
a few steps away, even from the highest-lying minimum. Moreover, the
betweenness centrality is largest, by orders of magnitude,
for the global minimum, so the majority of shortest paths
between all pairs of minima pass through 
the global minimum, i.e.~the global minimum is the central node of these networks. 
Hence finding the global minimum and
exploiting the funnelled/small-world structure of the landscape, 
is relatively straightforward from a numerical optimisation point of view.
Interestingly, though our results suggest that finding the global minimum of
the Thomson Problem may be relatively easy,
finding an answer to the Smale problem is NP-hard \cite{beltran2013harmonic}.
Constructing and analysing network of minima for the logarithmic version of the potential \cite{smale1998mathematical-2} 
may provide more concrete details on the mathematics behind Smale's 7th problem.

The present results agree with the observation \cite{Doye02} that funnelled energy 
landscapes display small world characteristics. 
In the future we will explore whether these two features are generally correlated,
or if a counterexample can be found. Another avenue for future work is to
extend our analysis to larger
particle numbers. Our previous work on the Thomson Problem
\cite{wales2006structure} found that global minima for
for $N > 400$ start to display alternative defect motifs. We
expect the potential energy 
landscape to display multiple funnels in this regime, and it will be
interesting to see whether the small world 
phenomenon found here will be preserved. 

In the future, we will further analyse the network properties of this model by including
weights and directions for the edges, depending on the 
barrier heights and kinetic transition rates. We also plan to 
develop more specific algorithms
to locate local and global minima by exploiting small-world properties \cite{mehta2014communication,mehta2015exploring,chen2015index}.
Analysing the appropriately weighted and directed networks of \textit{free}
energy minima \cite{rao2004protein} and transition states may provide additional
insight into the Thomson Problem. Furthermore, invesetigating network properties of higher index 
saddles may also provide insights to develop novel optimization algorithms  \cite{hughes2014inversion}.

\section{Acknowledgement}
DM was supported by a Australian Research Council DECRA fellowship no. DE140100867.
DZC was supported in part by the NSF under Grant CCF-1217906. We thank Carlos Beltr\'an and Edward Saff for their helpful remarks
the authors of \cite{Calef:2015} for clarifying their results.

\bibliographystyle{jcp}
\renewcommand{\bibname}{}
\bibliography{../bib/protein,../bib/wales,../bib/trapped,rosette,bibliography_NPHC_NAG,invert,NSF-2014,nh}
\end{document}